# Magnetic radial vortex stabilization and efficient manipulation driven by Dzyaloshinskii–Moriya interaction and spin-transfer torque

G. Siracusano[1], R. Tomasello[2], A. Giordano[1], V. Puliafito[3], B. Azzerboni[3], O. Ozatay[4], M. Carpentieri[5], G. Finocchio[1]

[1] Department of Mathematical and Computer Sciences, Physical Sciences and Earth Sciences, University of Messina, I-98166, Messina, Italy

[2] Department of Engineering, Polo Scientifico Didattico di Terni, University of Perugia, Terni, TR, I-50100 Italy

[3] Department of Engineering, University of Messina, I-98166, Messina, Italy

[4] Department of Physics, Bogazici University, 34342 Bebek/Istanbul, Turkey

[5] Department of Electrical and Information Engineering, Politecnico di Bari, via E. Orabona 4, I-70125 Bari, Italy

**Abstract**

Solitons are very promising for the design of the next generation of ultralow power devices for storage and computation. The key ingredient to achieve this goal is the fundamental understanding of their stabilization and manipulation. Here, we show how the interfacial Dzyaloshinskii–Moriya Interaction ($i$-DMI) is able to lift the energy degeneracy of a magnetic vortex state by stabilizing a topological soliton with radial chirality, hereafter called radial vortex. It has a non-integer skyrmion number $S$ ($0.5<|S|<1$) due to both the vortex core polarity and the magnetization tilting induced by the $i$-DMI boundary conditions. Micromagnetic simulations predict that a magnetoresistive memory based on the radial vortex state in both free and polarizer layers can be efficiently switched by a threshold current density smaller than $10^6$ A/cm$^2$. The switching processes occur via the nucleation of topologically connected vortices and vortex-antivortex pairs, followed by spin-wave emissions due to vortex-antivortex annihilations.

Magnetic solitons, such as domain walls (DWs) [1,2,3,4], vortices [5,6,7,8,9,10,11] and skyrmions [12,13,14,15,16,17], are attracting a growing interest for their potential technological applications in ultralow power devices beyond CMOS, thanks to their fundamental properties and the possible use of spin-transfer [18] and spin-orbit [2,3,4] torques for their nucleation and manipulation. For example, DWs and skyrmions are at the basis of racetrack memories [3,4,12,19], while vortices can be used as an alternative to the uniform state to store information in magnetoresistive devices [5,11], such as magnetic tunnel junctions (MTJs). Furthermore, the presence of the additional degree of freedom of the exchange-like interfacial Dzyaloshinskii–Moriya Interaction ($i$-DMI) [20] in the energy landscape of ferromagnetic materials has given rise to the possibility to speed-up DW motion in racetrack memories [3,4], and stabilize skyrmions in ultrathin ferromagnets [12,16]. While numerous theoretical and experimental studies have concentrated on the effect of the $i$-DMI on the static and dynamical characterization of DWs [3] and skyrmions [12,14,19], the understanding of its effect on vortices is still missing. In the absence of $i$-DMI, vortices are characterized by polarity and chirality. The former indicates the direction of the vortex core, +1 or -1 if the core points along the positive or negative out-of-plane direction. The latter is linked to the orientation of the in-plane magnetization around the core. In this letter, we show that a large enough $i$-DMI can stabilize a vortex state characterized by a radial chirality [21], hereafter called *radial vortex*, where the in-plane components of the magnetization point radially from the vortex core to the boundary (outward) or *vice versa* (inward), depending on vortex polarity and $i$-DMI sign. Fig. 1 shows the spatial distribution of the magnetization for different types of vortex. The chirality can be counterclockwise (CCW), clockwise (CW) and radial. Here, we use the term *circular vortex* to identify a vortex with CW or CCW chirality. For a fixed polarity, in a circular vortex, both CCW and CW configurations are energetically equivalent (degeneracy in the energy landscape) [9]. On the other hand, for a radial vortex, the core polarity fixes the chirality (radial and antiradial for positive or negative vortex core polarity, respectively), because of the non-symmetric $i$-DMI field. In other words, a large enough $i$-DMI removes the energy degeneracy of a vortex state.

From a fundamental point of view, the stabilization of a radial vortex gives rise to the possibility to create current densities with radial polarization for particle-trapping applications, such as skyrmion[22], analogously to what radially polarized beams can do in many optical systems [23].

In order to show the main features and the possible applications of the radial vortex, extensive micromagnetic simulations have been performed. The first part of this letter focuses on the fundamental properties of the radial vortex, in terms of stability as a function of the $i$-DMI, topology, nucleation as well as response to an in-plane field. The second part is dedicated to the analysis of the switching process of the radial vortex, underlining the differences with the circular vortex.

We consider a CoFeB disk having a diameter $d$=250nm and a thickness $t$=1.0nm to have an in-plane easy axis at zero external field. To add the *i*-DMI, we consider the CoFeB coupled with a Pt (heavy metal) layer as already experimentally observed [24]. The study is carried out by means of a *state-of-the-art* micromagnetic solver which numerically integrates the Landau-Lifshitz-Gilbert (LLG) equation [25,26,27], that includes the *i*-DMI contribution $\mathbf{h}_{i-\text{DMI}}$ [28,29,30]. Here, we show results achieved for different values of negative *i*-DMI parameter $D$ (in the rest of the letter we indicate always the magnitude $|D|$ of the *i*-DMI parameter), but both qualitatively and quantitatively similar results have been obtained for positive $D$ with the difference that the radial chirality is reversed for a fixed polarity.

*Stability.* The magnetization stability has been studied as a function of $|D|$ from 0.0 to 3.0 mJ/m$^2$ with a resolution of 0.1mJ/m$^2$. For each $|D|$, we have performed simulations for three different initial magnetization configurations (uniform state, radial and circular vortex). Fig.2(a) summarizes the final states achieved without thermal effects. Those magnetization patterns are independent of the initial state except in the range $0.7 \leq |D| \leq 1.2$mJ/m$^2$, where both uniform state and circular vortex are *minima*. Fig.3 shows the skyrmion number $S$ as a function of the *i*-DMI as computed from the topological density $S = \frac{1}{4\pi} \int \mathbf{m} \cdot \left( \partial_x \mathbf{m} \times \partial_y \mathbf{m} \right) dxdy$ [31]. At low *i*-DMI ($0.0 \leq |D| \leq 1.1$mJ/m$^2$), $S$ slightly increases from the zero *i*-DMI case ($|S|$=0.5) because of the *i*-DMI boundary conditions. As $|D|$ increases, the uniform state is stable in the region $0.7 \leq |D| \leq 1.6$mJ/m$^2$ ($|S| \approx 0$), while the radial vortex, characterized by $0.8 \leq |S| < 0.9$, is the final state in the range $1.7 \leq |D| \leq 2.3$mJ/m$^2$. This stability diagram can be qualitatively understood as follows. For in-plane materials the effect of the *i*-DMI can be seen as an additional shape anisotropy contribution [24], thus the different magnetic states (Fig.2(a)) can be linked to a trade-off among magnetostatic, exchange, and *i*-DMI energies. In particular, at low $|D|$, the magnetostatic energy is predominant, stabilizing a vortex with a circular chirality. Increasing $|D|$, the exchange ($0.7 \leq |D| \leq 1.6$mJ/m$^2$) and the *i*-DMI ($1.7 \leq |D| \leq 2.3$mJ/m$^2$) energies become the main contributions, leading to uniform and radial vortex states, respectively (Fig.3). We wish to stress that the $S$ computed by considering the core (region where the normalized *z*-component of the magnetization is larger than 0.1) of the radial vortex only is $|S|$=0.5 and does not depend on $|D|$. The additional contribution to $S$ is due to the *i*-DMI boundary conditions, which lead the magnetization near the sample edges to be tilted out-of-plane [24].

For $|D|\geq 2.4$mJ/m$^2$, the compensation of the in-plane anisotropy favors the out-of-plane easy axis of the magnetization, and multi-domain patterns arise, because the *i*-DMI energy is minimized as the number of Néel DWs increases. For example, when $|D|=2.5$mJ/m$^2$, a "horseshoe"-like domain is obtained, with internal magnetization aligned along the positive *z*-direction, separated from an opposite domain through Néel DWs. When $|D|=3.0$mJ/m$^2$, a number of Néel DWs are nucleated, leading to a "spider web"-like domains. The equilibrium states are in agreement with our computations based on energetic arguments.

Fig. 2(b) shows the final states (after 500ns) achieved in the presence of thermal fluctuations at room temperature (*T*=300 K). By comparing Fig.2(a) and 2(b), it can be observed how the thermal effects influence the stability. In particular, our results show that the stability for the radial vortex is a remanent state at room temperature in the range $1.8\leq|D|\leq 2.1$mJ/m$^2$. Similar results have been achieved for disk diameters from 100nm to 350nm [30] (see Supplemental Material) after the study at 0K.

*Nucleation.* A simple procedure to nucleate a radial vortex is a relaxation process from an out-of-plane saturation state. Firstly, a large external out-of-plane magnetic field of 500 mT is applied to saturate the magnetization. The field is then quasi-statically decreased, and, below a field value of about 100 mT up to 0 mT, the *i*-DMI promotes the formation of the radial vortex [30] (see Supplemental Material MOVIE 1 for the nucleation process achieved at $|D|=2.0$mJ/m$^2$). This nucleation procedure is robust to a possible tilting angle of the external field with respect to the out-of-plane direction up to +/-15 degree, and to thermal fluctuations at room temperature [30]. Similar results are obtained for *i*-DMI values in the range $1.7\leq|D|\leq 2.3$mJ/m$^2$.

*Response to in-plane external field.* An in-plane field ($H_{ext}$) shifts circular and radial vortex differently. Let consider a field applied along the positive *x*-direction. Fig. 4(a) displays the results for the radial vortex ($|D|=2.0$mJ/m$^2$). The external field favors the expansion of the region where the *x*-component of the magnetization is parallel to the external field, thus inducing a shift of the core along the negative *x*-direction (insets of Fig.4(a)). An $H_{ext}=5.0$mT yields a uniform magnetic state. Fig.4(b) shows the results for the circular vortex ($|D|=0.0$mJ/m$^2$). The displacement occurs along the direction perpendicular to the external field (insets Fig.4(b)) [7,32,33,34,35]. Together with the qualitatively different vortex core shift, the field value that leads to the radial vortex core expulsion is twice the one of the circular vortex. This is ascribed to the *i*-DMI and, in particular, to its boundary conditions, in fact, the expulsion field increases as a function of $|D|$. The key reason for that is the need of a larger field to align the out-of-plane tilted spins at the sample edges perpendicular to the vortex core motion. We stress the fact that the different behavior exhibited by the two types of vortices can be considered as a "finger print" to detect them in experimental measurements by imaging the vortex core shifting.

*Switching.* The second part of this work deals with the switching of radial vortex, implying its possible use in magnetoresistive memories. We focus on the radial vortex at $|D|=2.0\text{mJ/m}^2$ [24,36,37,38,39]. The device is a circularly shaped MTJ, where a CoFeB polarizer is also coupled to a Pt layer. Geometrical and physical parameters of the free layer are the same of the disk described above. The polarizer is designed to have a fixed radial vortex with positive polarity as ground state. A possible strategy to have the polarizer less sensitive to the spin-transfer-torque is tailoring the percentage of the Co, Fe and B to increase the saturation magnetization. A current density $J_{MTJ}$ flows perpendicularly through the whole free layer cross section [30].

It is possible to identify two switching mechanisms occurring at low ($J_{MTJ}<11\text{MA/cm}^2$) and high current densities ($J_{MTJ}\geq11\text{MA/cm}^2$), respectively. We describe the results when the initial state is the radial vortex (snapshot Fig.5(a)) only, but qualitatively similar mechanisms occur for antiradial → radial switching.

The mechanism at low current density is characterized by a complex process, involving radial and antiradial vortices, antivortices, topologically connected radial vortices, edge radial and antiradial vortices [40]. It starts with the motion of the radial vortex core towards the boundary of the free layer, accompanied by a core expansion, which gives rise to topologically connected radial vortices, and by a nucleation of an edge antiradial vortex (snapshot Fig.5(b)). Afterwards, the topologically connected radial vortices split into two antivortices with positive polarity and an edge radial vortex, while two antiradial vortices and two edge antiradial vortices are nucleated as well (snapshot Fig.5(c)). Subsequently, the edge radial vortex annihilates, one antiradial vortex-antivortex pair collapses emitting a spin wave, and the edge antiradial vortex turns into an antiradial vortex. Therefore, before the accomplishment of the switching, a particular magnetization configuration is obtained, composed of two antiradial vortex–antivortex pairs and one antiradial vortex (snapshot Fig.5(d)). Finally, the two antiradial vortex–antivortex pairs collapse emitting spin-waves, and only the antiradial vortex remains at the center (snapshot Fig.5(e)) marking the end of the switching (see Supplemental Material MOVIE 2 for the switching process achieved at $J_{MTJ}=5\text{MA/cm}^2$).

At high currents ($>11\text{MA/cm}^2$), the second switching mechanism takes place. The application of the current destroys the radial vortex and promotes the formation of an antiradial vortex together with an antiradial vortex-antivortex pair. Subsequently, the pair collapses and only the antiradial vortex remains, accomplishing the switching (see Supplemental Material MOVIE 3 for the switching process achieved at $J_{MTJ}$=15MA/cm$^2$).

For the sake of clarity, we have compared the radial and circular vortex (|$D$|=0.0mJ/m$^2$) switching. The first difference is that, for a circular vortex, the switching mechanism is unique. Starting from a ground state (positive polarity and CW chirality), it involves a core polarity reversal followed by a spin-wave excitation, and, afterwards the chirality switching. The final state is then a CCW circular vortex with negative polarity (see Supplemental Material MOVIE 4 for an example of the switching process at $J_{MTJ}$ =5MA/cm$^2$). This switching process is different from previous mechanisms reported in literature involving edge solitons [40], gyrotropic motion [41], or vortex-antivortex pairs [42].

Fig. 6 summarizes the switching time $t_s$ as a function of the current density for the circular and radial vortex. The key result is the achievement of a critical current $J_{MTJ}$=0.4MA/cm$^2$ for the radial vortex ($T$=0K), one order of magnitude lower than the one of the circular vortex ($J_{MTJ}$=4MA/cm$^2$). However, for $J_{MTJ} \geq 4.0$MA/cm$^2$, the switching time is shorter for the circular vortex, being its switching mechanism less complex. For the radial vortex, the thermal fluctuations at room temperature ($T$=300K) reduce the critical current density to 0.1MA/cm$^2$ (Ikeda *et al.* [43] and Amiri *et al.* [44] measured 2MA/cm$^2$, whereas Mangin *et al.* 7MA/cm$^2$ [45]), while keeping the same value for the circular vortex. Moreover, the radial vortex switching time decreases below the one of the circular vortex, because the thermal fluctuations delete the nucleation of most of the metastable states observed during the radial vortex switching mechanisms at $T$=0K. The inset shows an example of the statistical distribution of the switching time for $J_{MTJ}$=0.5MA/cm$^2$, as obtained by performing 100 realizations at $T$=300K. We represent the close qualitatively agreement between the histogram (cyan bars) and the corresponding Gaussian distribution (red curve) of the data, as built using the mean ($\mu$=23.3ns) and standard deviation ($\sigma$=3.5ns) of the numerical results.

A final remark is about the effect of the damping on the critical current density: our simulations show a value <0.55MA/cm$^2$ at $\alpha$ =0.1.

In summary, we have found that a large enough *i*-DMI lifts the chirality degeneracy of a circular vortex state, promoting the formation of a vortex with a radial chirality. The results have shown that this topological state is robust against thermal fluctuations and can be used for room temperature applications. We have suggested a way for the experimental detection of the radial vortex by the application of an in-plane magnetic field. Besides, the radial vortex switching exhibits very low critical current densities, one order of magnitude lower than the circular vortex state, which are suitable for the development of low power storage devices and non-uniform magnetization state based magnetoresistive memories. Finally, the switching process gives rise to an intrinsic reversal of both polarity and radial chirality of the vortex because of the topological constrain introduced by the *i*-DMI, with the relevant advantage to achieve a full TMR signal change in the characterization of parallel and anti-parallel states.

*Acknowledgments.* G. S. and R. T. contributed equally. This work was supported by the project PRIN2010ECA8P3 from Italian MIUR and by the bilateral agreement Italy-Turkey (TUBITAK-CNR) project (CNR Grant #: B52I14002910005, TUBITAK Grant # 113F378) "Nanoscale magnetic devices based on the coupling of Spintronics and Spinorbitronics". R. T. acknowledges Fondazione Carit - Projects – "Sistemi Phased-Array Ultrasonori", and "Sensori Spintronici". The simulations were performed in the Scientific Computing Laboratory (SCL) of the University of Messina, Italy.

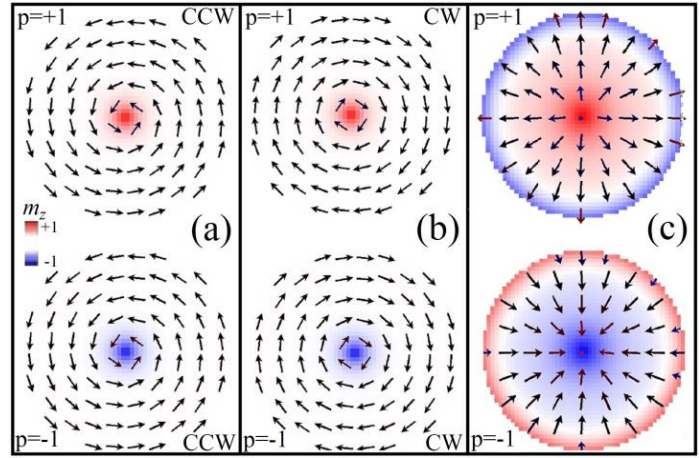

FIG. 1: Spatial distribution of magnetization for different types of vortex with positive (top) and negative (bottom) polarity. (a) Counterclockwise (CCW) circular vortex. (b) Clockwise (CW) circular vortex. (c) Radial vortex. A color scale linked to the out-of-plane component of the magnetization is also shown (red positive, blue negative).

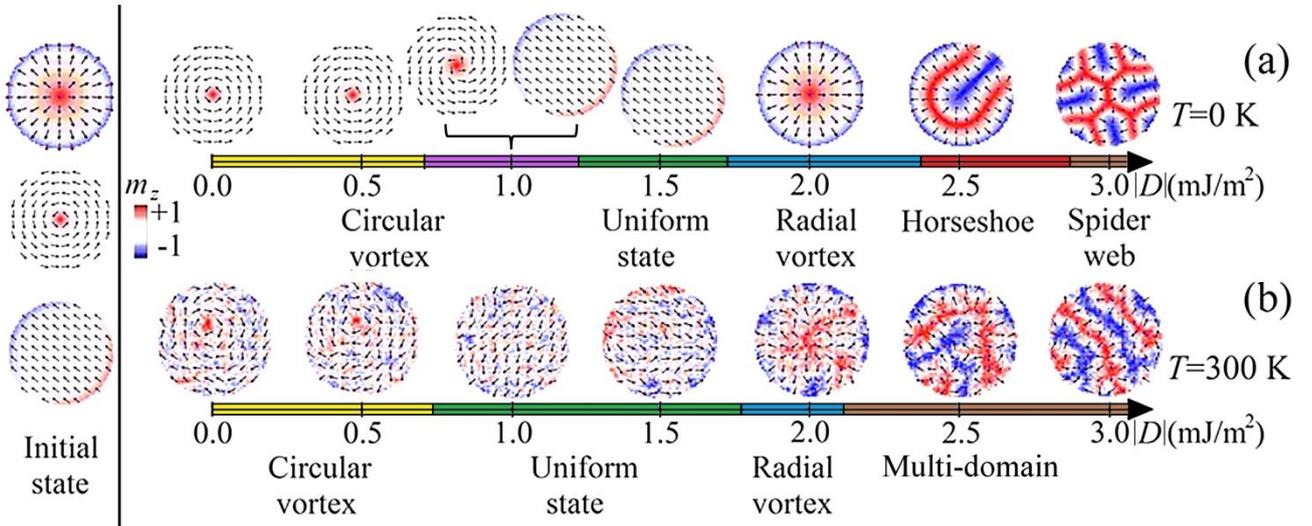

FIG. 2: Equilibrium configurations of the magnetization as a function of $|D|$ at zero external field when (a)$T$=0K, and (b)$T$=300K. The snapshots are calculated for $|D|$=0.0, 0.5, 1.0, 1.5, 2.0, 2.5, and 3.0 mJ/m$^2$. For $|D|$=1.0mJ/m$^2$ in (a), the two represented magnetization configurations depend on the initial state. The colors refer to the $z$-component of the magnetization (blue negative, red positive). Left inset: the initial state can be radial or circular vortex, as well as the uniform state.

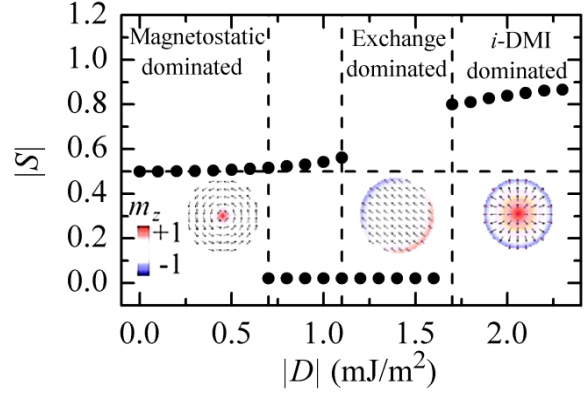

FIG. 3: Skyrmion number as a function of $|D|$. Four different regions are indicated and separated by vertical dashed lines. Insets: examples of spatial distribution of the magnetization related to each region. The colors refer to the $z$-component of the magnetization (blue negative, red positive).

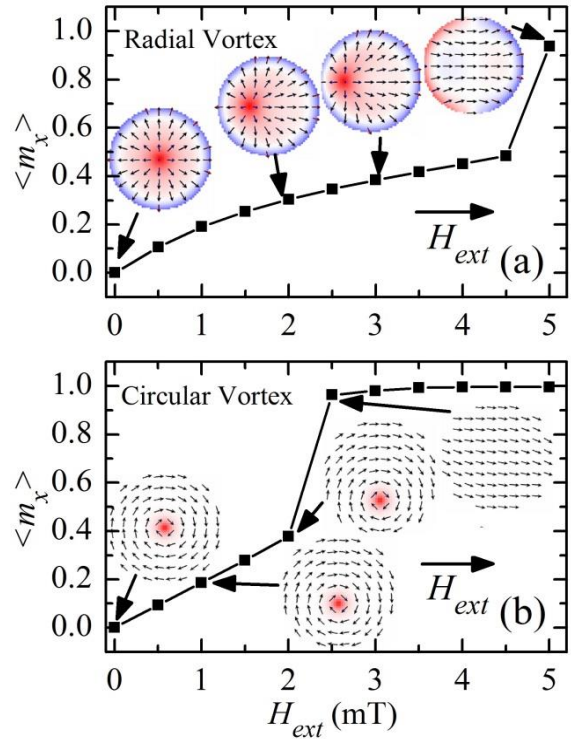

FIG. 4: $x$-component of the normalized magnetization as a function of the external field applied along the positive $x$-direction for (a) the radial vortex, when $|D|=2.0\text{mJ/m}^2$, and (b) the circular vortex, when $|D|=0.0\text{mJ/m}^2$. Insets: spatial distributions of the magnetization where the colors indicate the $z$-component (blue negative, red positive). The snapshots for different fields are represented, and the direction of the applied in-plane field is illustrated.

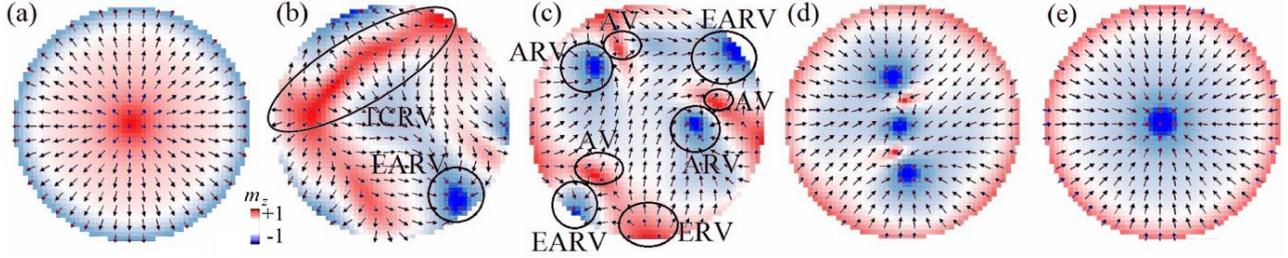

FIG. 5: Example of snapshots for the radial vortex switching at $J_{MTJ}$=5 MA/cm$^2$. (a) Initial state: radial vortex. (b) Topologically connected radial vortices together with an edge antiradial vortex. (c) Three antiradial vortex–antivortex pairs, together with an edge radial and antiradial vortex. (d) Configuration before the accomplishment of the switching: two antiradial vortex-antivortex pairs and one antiradial vortex are clearly observed. (e) Final state: antiradial vortex. The colors are linked to the *z*-component of the magnetization (red positive, blue negative). The acronyms TCRV, ARV, RV, AV, ERV, and EARV indicate the topologically connected radial vortices, antiradial vortex, radial vortex, antivortex, edge radial vortex, and edge antiradial vortex, respectively.

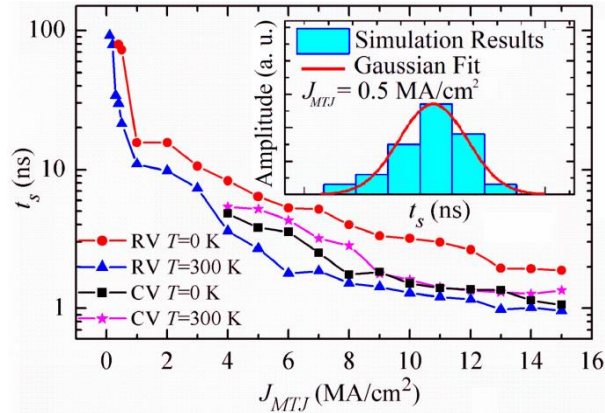

FIG. 6: The main panel shows the switching time as a function of the current density for the radial vortex (RV) and the circular vortex (CV) at *T*=0K and *T*=300K, as indicated in the legend. Inset: example of histogram related to 100 realizations performed for RV at *T*=300K for $J_{MTJ}$=0.5MA/cm$^2$, and the corresponding Gaussian fit.

# SUPPLEMENTAL INFORMATION
## Supplementary note 1

*Micromagnetic model.* The first part of the study is carried out by means of a *state-of-the-art* micromagnetic solver [1,2] which numerically integrates the Landau-Lifshitz-Gilbert (LLG) equation by applying the time solver scheme Adams-Bashforth with a second order Adams-Moulton procedure as a corrector (AB3M2 [2]):

$$(1+\alpha^2)\frac{d\mathbf{m}}{d\tau} = -(\mathbf{m}\times\mathbf{h}_{\text{eff}}) - \alpha\mathbf{m}\times(\mathbf{m}\times\mathbf{h}_{\text{eff}}) \quad (1.1)$$

where $\alpha$ is the Gilbert damping, $\mathbf{m}$ is the normalized magnetization, and $\tau = \gamma_0 M_s t$ is the dimensionless time, with $\gamma_0$ being the gyromagnetic ratio, and $M_s$ the saturation magnetization. $\mathbf{h}_{\text{eff}}$ is the normalized effective field, which includes the exchange, magnetostatic, anisotropy and external fields, as well as the *i*-DMI and the thermal field. The *i*-DMI contribution $\mathbf{h}_{i-\text{DMI}}$ is derived from the functional derivative of the *i*-DMI energy density $\varepsilon_{i-\text{DMI}} = D[m_z \nabla\cdot\mathbf{m} - (\mathbf{m}\cdot\nabla)m_z]$ under the hypothesis of ultrathin film $\left(\frac{\partial \mathbf{m}}{\partial z} = 0\right)$ [3]:

$$\mathbf{h}_{i-\text{DMI}} = -\frac{1}{\mu_0 M_S}\frac{\delta\varepsilon_{i-\text{DMI}}}{\partial\mathbf{m}} = -\frac{2D}{\mu_0 M_S}\left[(\nabla\cdot\mathbf{m})\hat{z} - \nabla m_z\right] \quad (1.2)$$

where $D$ is the parameter taking into account the intensity of the *i*-DMI, $m_z$ is the out-of-plane component of the normalized magnetization, $\mu_0$ is the vacuum permeability, and $\hat{z}$ is the unit vector along the out-of-plane direction. The *i*-DMI affects the boundary conditions of the ferromagnetic sample in the following way $\frac{d\mathbf{m}}{dn} = \frac{1}{\xi}(\hat{z}\times\mathbf{n})\times\mathbf{m}$, where $\mathbf{n}$ is the unit vector normal to the edge, $\xi = \frac{2A}{D}$ ($A$ is the exchange constant) is a characteristic length in presence of the *i*-DMI [4]. The thermal effect is considered as an additional stochastic term $\mathbf{h}_{\text{th}}$ to the deterministic effective field in each computational cell $\mathbf{h}_{\text{th}} = (\boldsymbol{\chi}/M_S)\sqrt{2(\alpha K_B T / \mu_0 \gamma_0 \Delta V M_s \Delta t)}$, where $K_B$ is the Boltzmann constant, $\Delta V$ is the volume of the computational cubic cell, $\Delta t$ is the simulation time step, $T$ is the temperature of the sample, and $\boldsymbol{\chi}$ is a three-dimensional white Gaussian noise with zero mean and unit variance, and it is uncorrelated for each computational cell [5,6]. The discretization cell used is 5x5x1nm$^3$ benchmarked with simulations using discretization cell of 2.5x2.5x1nm$^3$ Except for the nucleation process, all the simulations have been performed at zero external field. The physical parameters used in this study are: $M_S$=1000kA/m, $A$=20pJ/m, perpendicular anisotropy constant $k_u$=0.50MJ/m$^3$, $\alpha$ =0.02 [7]. The results discussed in the main text are robust to the following range of parameters 0.01< $\alpha$ <0.1, 15pJ<$A$<25pJ, and 900kA/m<$M_S$<1100kA/m.

In order to study the switching dynamics in the second part of the work, the Eq. (1.1) has been extended to include Oersted field [1], as an additional contribution to the effective field, and the Slonczewski spin-transfer-torque term for the MTJs [8,9,10]:

$$-\frac{g|\mu_B|J_{MTJ}}{e\gamma_0 M_S^2 t}\varepsilon(\mathbf{m},\mathbf{m_p})\left[\mathbf{m}\times(\mathbf{m}\times\mathbf{m_P})-q(V)(\mathbf{m}\times\mathbf{m_P})\right] \quad (1.3)$$

where $g$ is the Landè factor, $\mu_B$ is the Bohr magneton, $e$ is the electron charge, $t$ is the thickness of the free layer. $J_{MTJ}$ is the current density flowing perpendicularly through the whole free layer cross section (nanopillar geometry), $\varepsilon(\mathbf{m},\mathbf{m_p}) = 2\eta\left(1+\eta^2\mathbf{m}\cdot\mathbf{m_p}\right)^{-1}$ characterizes the angular dependence of the Slonczewski spin torque term, where the spin-polarization factor $\eta$ is fixed to 0.66, $\mathbf{m}$ is the magnetization of the free layer, $\mathbf{m_P}$ is the magnetization of the pinned layer, and $q(V)$ is the voltage dependent parameter for the perpendicular torque [9]. The magnetization distribution of the polarizer is non-uniform, i.e. a radial vortex with positive polarity. In order to have the magnetization of the polarizer less sensitive to the STT, it is possible to increase its saturation magnetization, thickness or both. Simple estimations based on Eq. (1.3) show that a saturation magnetization of 1.3x10$^6$ A/m should be sufficient with the device parameter used for the free layer in this work. Fig. S1 represents a sketch of the analyzed MTJ, where a Cartesian coordinate system has been also introduced. In our computations, we have neglected the dipolar coupling between the two ferromagnets. This scenario can be created experimentally by considering two ferromagnets anti-ferromagnetically coupled via interlayer exchange coupling [11] and one of them coupled with the Pt in order to induce the *i*-DMI and stabilize the radial vortex.

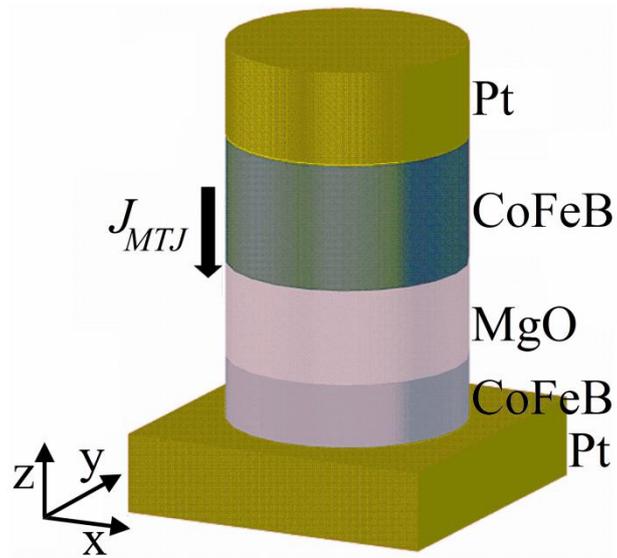

FIG. S1: Sketch of the MTJ under investigation. The CoFeB free layer and polarizer are coupled to the Pt heavy metal to originate the *i*-DMI in order to stabilize the radial vortex state.

**Supplementary note 2**

*Nucleation process.* We performed micromagnetic simulations at zero and room temperature $T=300$K, by applying an external out-of-plane field to nucleate the radial vortex, in the *i*-DMI region where the radial vortex is stable ($1.7 \leq |D| \leq 2.3$ mJ/m$^2$), and for the disk with $d=250$nm. We describe the results only for $|D|=2.0$ mJ/m$^2$, but similar outcomes are achieved for other *i*-DMI values. Firstly, we saturate the sample by means of an external field equal to 500mT. Afterwards, the sample is relaxed by quasi-statically decreasing the external field up to zero. Around a field value of 100mT, the radial vortex starts to be nucleated, both at zero temperature (Fig. S2(a)) and room temperature (Fig. S2(b)).

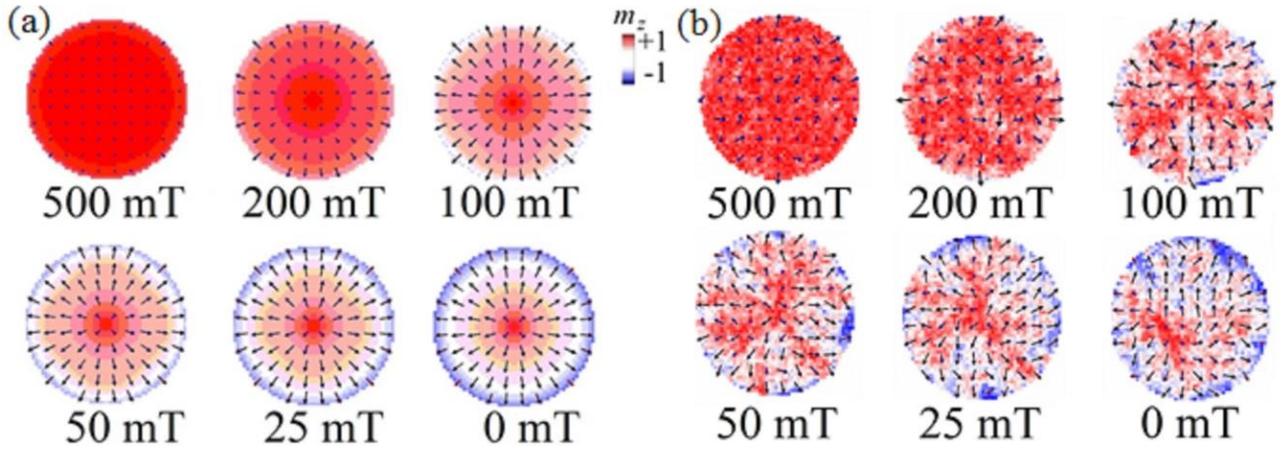

FIG. S2: Snapshots of the spatial distribution of the magnetization (disk with a diameter $d=250$ nm) for the nucleation process of the radial vortex achieved when $|D|=2.0$ mJ/m$^2$ at (a) $T=0$K and (b) $T=300$K. The external field is indicated below each snapshot. The color is related to the out-of-plane component of the magnetization (red positive, and blue negative).

**Supplementary note 3**

*Stability as a function of the sample size.* We performed micromagnetic simulations by considering different disk diameters $d$=50, 100, 150, 200, 300, 350, and 400nm, while keeping the thickness $t$=1.0nm. Fig. S3 summarizes the stability diagram as a function of the $i$-DMI parameter for the radial vortex. There exists an optimal value of the diameter ($d$=150nm) that maximizes the width of the radial vortex stability range. Moreover, for the disk diameters $d$=50 and 400nm, the radial vortex is no longer stable. In the former, the exchange energy is the dominant term, promoting the stabilization of the uniform state; in the latter, the larger magnetostatic energy favors the formation of multi-domain patterns at lower values of $|D|$.

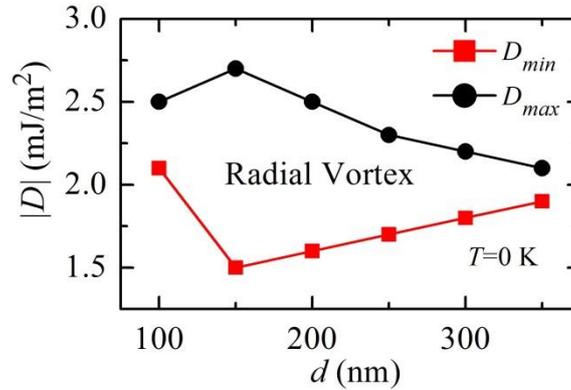

FIG. S3: $i$-DMI stability range for the radial vortex as a function of the diameter of the disk.

Fig. S4 depicts a detailed stability diagram for disk diameters $d$=150nm (Fig. S4(a) and (b)), and $d$=350nm (Fig. S4(c) and (d)), at zero and room temperature $T$=300K. Similar stability diagrams are achieved by changing the disk diameter, namely the circular and radial vortex, the uniform state, and multi-domain patterns are still observed, except for $d$=50 and 400nm, as discussed above. The important result is that, when decreasing the disk size, the circular vortex is no longer stable at room temperature, being replaced by the uniform state. This implies another advantage of the radial vortex on the circular one, i.e. smaller devices can be designed, leading to an increase of the memory density if the radial vortex is used as building block in soliton-based storage applications.

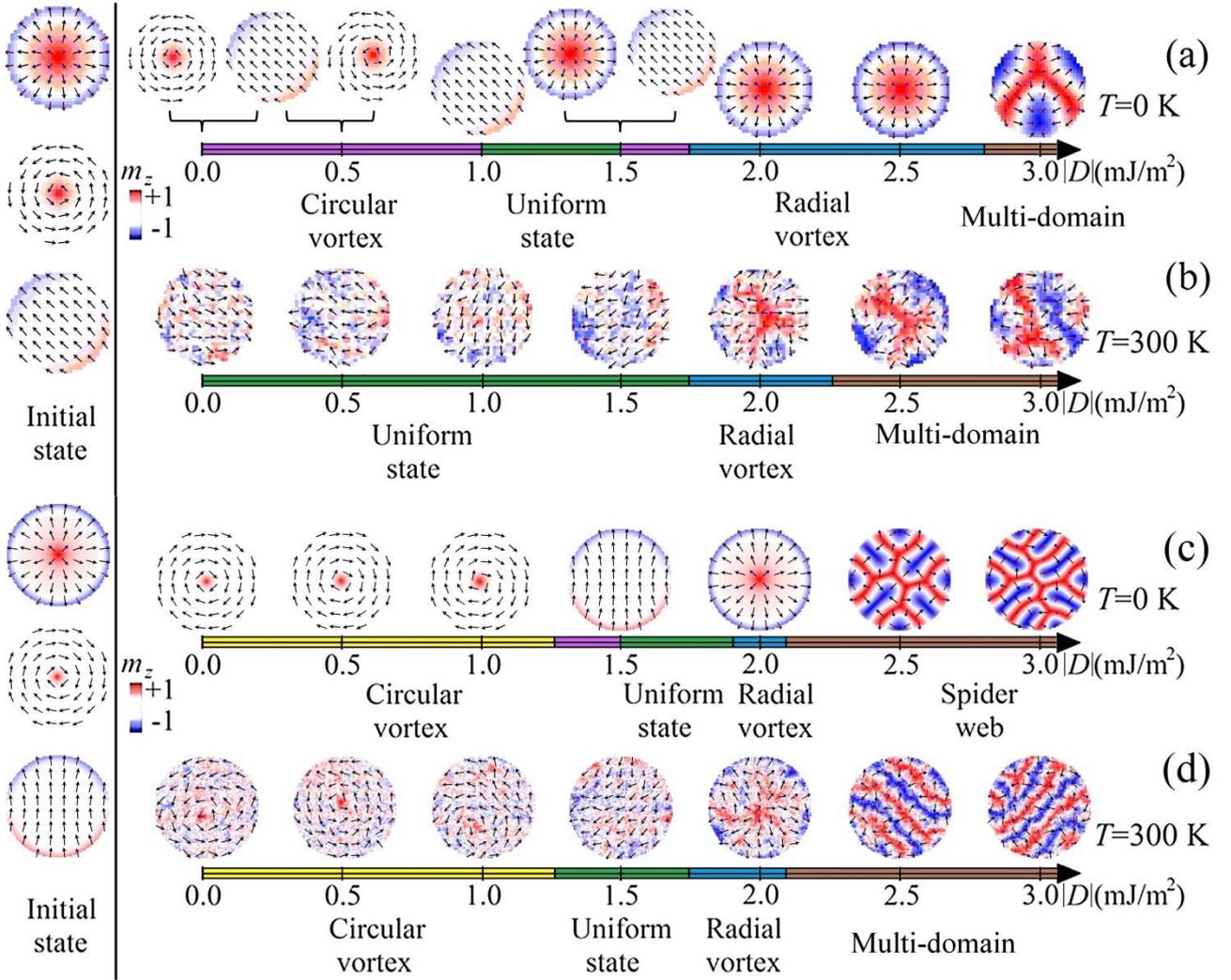

FIG. S4: Equilibrium configurations of the magnetization as a function of |D| at zero external field when (a) and (b) d=150nm, at T=0K, and T=300K, respectively, and (c) and (d) d=350nm, at T=0K, and T=300K, respectively. The snapshots are calculated for |D|=0.0, 0.5, 1.0, 1.5, 2.0, 2.5, and 3.0mJ/m$^2$. The colors refer to the z-component of the magnetization (blue negative, red positive). Left inset: the initial state can be radial or circular vortex, as well as the uniform state. The black curly brackets indicate the two stable magnetization configurations which depend on the initial state.

**Supplementary note 4**

*Comparison with skyrmions.* Recently, a magnetic soliton, i.e. skyrmion, has been attracting a lot of interest for its topological properties as well as for a wide range of potential applications. We wish to make a comparison between radial vortex and skyrmion in terms of stability, nucleation and switching processes, referring to our micromagnetic achievements for the former, and to the main results published in literature for the latter [12,13,14,15,16,17].

The main difference is that skyrmions are stabilized in out-of-plane materials and characterized by an *integer* skyrmion number, while radial vortices are stabilized in in-plane materials and have a *non-integer* skyrmion number. Another difference is that, as we can observe in Fig. S5, the spatial profiles of the out-of-plane component of magnetization $m_z$ of a radial vortex (a) and a skyrmion (b) exhibit a different shape.

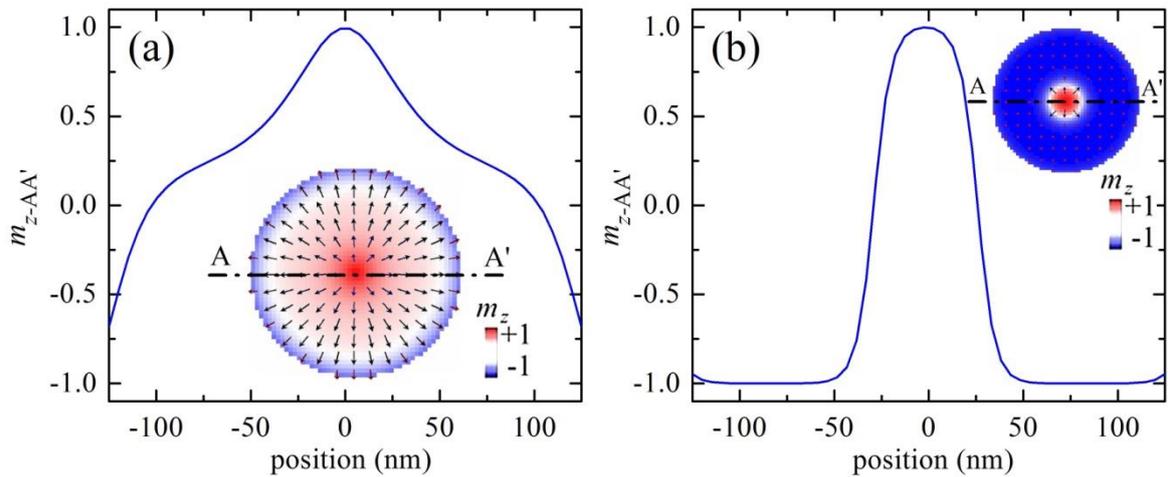

Fig. S5 - Spatial profile of *z*-component of the normalized magnetization corresponding to the section AA' as indicated in the insets for: (a) the radial vortex, and (b) a Nèel skyrmion, when $|D|=2.0$ mJ/m$^2$. Insets: spatial distributions of the magnetization $m_z$ where the colors indicate the *z*-component (blue negative, red positive).

Stability. To stabilize a skyrmion, two key ingredients are necessary: a large enough *i*-DMI for Néel skyrmion, and an out-of-plane easy axis of the magnetization. The latter can be achieved in two ways: (*i*) application of large out-of-plane fields (usually for bulk materials [12,14]), (*ii*) use of materials with high perpendicular anisotropy (typical of ultrathin ferromagnets coupled to heavy metals [15,16,18]). On the other hand, the radial vortex can be stabilized in materials with an in-plane easy axis, where, also in this case, the *i*-DMI plays a significant role.

Nucleation. Several procedures have been experimentally investigated to nucleate skyrmions. Single skyrmions have been nucleated near zero temperature by means of spin-polarized scanning tunneling microscope [13] in perpendicularly magnetized ultrathin films. In Ref. [16], skyrmions have been obtained through a conversion of domain walls in a symmetrically designed wire containing a geometrical constriction. A third procedure concerns relaxation processes by applying an out-of-plane field. In particular, the skyrmion nucleation is achieved by increasing the field starting from a multi-domain state, or by decreasing it, starting from the out-of-plane saturation state [17].

Numerical works have shown the possibility to nucleate a single skyrmion via different methods. Iwasaki *et al.* [14] simulated a notch in the perpendicularly magnetized wire, and, by applying an electric current pulse, Bloch skyrmions from the notch were nucleated via the spin-transfer torque. A similar shape-assisted process combined with a microwave field was used in [19]. In Ref. [15], a single Néel skyrmion has been nucleated by means of the spin-transfer torque from an out-of-plane spin-polarized current locally injected.

For the radial vortex, here we propose a simple procedure for its nucleation based on a relaxation process by applying an out-of-plane field, as done for skyrmions in [17].

<u>Switching.</u> This is a very interesting aspect for applications in data storages. The key difference between radial vortex and skyrmion is how the information is coded .

The bit "1"/"0" is coded in both polarity and chirality of a radial vortex. The switching is then related to a change of both polarity and chirality .

For skyrmions, the bit "1"/"0" is coded by the presence or absence of the skyrmion itself (e.g. racetrack memories [20]).The switching is then related to the nucleation and annihilation of a skyrmion.